\documentclass[revtex]{emulateapj}
\journalinfo{The Astrophysical Journal, in press}
\submitted{Accepted for publication in the Astrophysical Journal}

\usepackage{graphicx}

\shorttitle{Stacking Polarized Intensity}
\shortauthors{Stil et al.}

\begin{document}

\title{Degree of Polarization and Source Counts of Faint Radio Sources\\ from Stacking Polarized Intensity}

\author{J. M. Stil}
\affil{Department of Physics and Astronomy, University of Calgary, 2500 University Drive NW, Calgary AB T2N 1N4, Canada}
\author{B. W. Keller} 
\affil{Department of Physics and Astronomy, McMaster University, 1280 Main Street W, Hamilton ON, L8S 4M1, Canada}
\author{S. J. George}
\affil{Department of Physics and Astronomy, University of Calgary, 2500 University Drive NW, Calgary AB T2N 1N4, Canada}
\author{A. R. Taylor}
\affil{Department of Astronomy, University of Cape Town \& Physics Department, University of Western Cape\\ Private Bag X3, Rondebosch 7701, Republic of South Africa\\}

\begin{abstract}
We present stacking polarized intensity as a means to study the
polarization of sources that are too faint to be detected individually
in surveys of polarized radio sources. Stacking offers not only high
sensitivity to the median signal of a class of radio sources, but also
avoids a detection threshold in polarized intensity, and therefore an
arbitrary exclusion of source with a low percentage of
polarization. Correction for polarization bias is done through a Monte
Carlo analysis and tested on a simulated survey. We show that the
non-linear relation between the real polarized signal and the detected
signal requires knowledge of the shape of the distribution of
fractional polarization, which we constrain using the ratio of the
upper quartile to the lower quartile of the distribution of stacked
polarized intensities. Stacking polarized intensity for NVSS sources
down to the detection limit in Stokes I, we find a gradual increase in
median fractional polarization that is consistent with a trend that
was noticed before for bright NVSS sources, but is much more gradual
than found by previous deep surveys of radio
polarization. Consequently, the polarized radio source counts derived from our
stacking experiment predict fewer polarized radio sources for future
surveys with the Square Kilometre Array and its pathfinders.
\end{abstract}


\section{Introduction}

The degree of polarization of faint radio sources is of astrophysical
interest because it measures the regularity of magnetic fields in
these radio sources. While a significant fraction of radio sources
brighter than 100 mJy are luminous sources, brighter than the
luminosity boundary between FRI and FRII radio galaxies, the radio
source population around 30 mJy is dominated by radio galaxies below
the FRI/FRII luminosity boundary. If this gradual transition to less
luminous radio galaxies is accompanied by a difference in structure,
as in the case of FRI/FRII radio galaxies, it is conceivable that the
polarization properties of faint radio sources is different from those
of bright radio sources. Around a flux density of 1 mJy, an
increasing fraction of radio sources is powered by star formation
instead of an active galactic nucleus (AGN), with a potentially
significant contribution of radio-quiet QSOs. The polarization
properties of these objects may also be quite different from the
bright radio source population studied so far.

Characterization of the polarization of radio sources is difficult
because the polarized signal is only a few percent of the total
brightness of the source. The NRAO VLA Sky Survey
\citep[NVSS,][]{condon1998} has detected 1.8 million sources in total
intensity, but only 14\% of these have a reported (raw) peak polarized
signal greater than $3\sigma$. \citet{mesa2002} and \citet{tucci2004}
found that the median percentage polarization of steep-spectrum NVSS
sources brighter than 100 mJy increase from $1\%$ for sources brighter
than 800 mJy to $2\%$ for sources between 100 mJy and 200 mJy. For
fainter sources, the median polarized intensity of $2\%$ would be
below the formal detection limit. Fainter polarized sources exist in
the NVSS, but these are highly-polarized sources in the tail of the
distribution. The application of the analysis presented in this paper
therefore begins at the $\sim$100 mJy level in total flux density,
even though a substantial fraction of the sources is still detected in
polarization.

The polarization of radio sources fainter than $100$ mJy at 1.4 GHz
has been studied directly in deep fields with a high sensitivity but
small survey area
\citep[e.g.][]{taylor2007,grant2010,subrahmanyan2010} with sensitivity
around $50\ \rm \mu Jy$, and a smaller field by \citet{rudnick2013} at
$3\ \rm \mu Jy$. \citet{macquart2012} applied forward fitting of
polarized source counts for a wider but shallower ($89\ \rm \mu Jy$)
survey to investigate the polarization of sources below the formal
detection threshold. The deep fields currently provide the
polarization data for sources fainter than $\sim 100$ mJy.  The long
integration times required to detect faint sources in polarization
limit the survey area of these fields, and therefore the attainable
sample size.  Published deep surveys were made with different
telescopes and apply different methods for detection, which raises
complications for combining results from different deep surveys. With
a sample size of the order of 100 detections in polarization, these
deep fields cannot constrain the polarization of relatively rare
objects such as the flat spectrum radio sources considered by
\citet{mesa2002} and \citet{tucci2004}. This limitation is a strong
driver for the large collecting area and large instantaneous bandwidth
of the international Square Kilometre Array (SKA) telescope
(\url{http://skatelescope.org}).  In anticipation of future wide and
deep polarization surveys, we explore stacking polarized intensity to
derive polarization properties of sources well below the detection
limit of current surveys.

Stacking is a statistical approach to derive the mean or median flux
density of a class of sources that cannot be detected individually in
a survey. If the position of a sample of sources is known, the
intensities at the recorded positions can be combined by taking the
average or the median. In this paper we focus on stacking radio
surveys, in particular polarized intensity from the NVSS. Stacking was
first applied to radio data by \citet{white2007}, addressing the
radio-loud/radio-quiet dichotomy of active galactic nuclei with
optically selected samples from the SDSS, and radio data from the
FIRST survey \citep{white2007}. The technique has also been applied to
study the infrared-radio correlation of galaxies by stacking the radio
emission of galaxies with nearly the same infrared flux density
\citep{garn2009,jarvis2010}. Confusion is a potential problem for
stacking if the source density in the reference catalogue leaves less
than several beams per source in the stacked survey.  The
effectiveness of stacking depends further on completeness and
astrometric accuracy of the input source catalog, and the degree to
which the stacked survey may resolve a significant fraction of the
sample. In this paper we will discuss some specific challenges related
to stacking polarized intensity.

Stacking polarization is attractive because the polarized signal is
intrinsically weak, and the total-intensity source catalogue is
available to define source positions. Stacking polarized intensity is
therefore unique in the sense that the reference sample is derived
from the same survey - in total intensity.  We have performed
experiments involving other radio surveys but focus here on stacking
of linear polarization from the NVSS for subsamples of the NVSS source
catalogue. We also limit the analysis to stacking polarized intensity,
but we point out that stacking Stokes parameters $Q$ and $U$ individually
can be meaningful when testing for the alignment of linear
polarization with source morphology, and that stacking circular
polarization is another application of the procedure outlined in this
paper, provided that the absolute value of Stokes V is stacked.

If polarized intensity is stacked for sources in a narrow range of
total flux density, the derived median polarized intensity yields a
median percentage polarization for that sample.  Stacking linear
polarization therefore allows us to investigate the fractional
polarization of radio sources as a function of flux density.  This in
turn can be combined with total source counts to derive polarized
radio source counts to very low flux densities, as a predictor for the
number density of rotation measures produced by future deep
polarization surveys with the SKA, as done
previously by \citet{bg04}, \citet{osullivan2008}, and
\citet{stil2009a}. Fitting deep polarized source counts and
total-intensity source counts simultaneously with models of the
cosmological evolution of radio sources provides information on the
relation between the cosmic evolution of magnetic field properties and
radio source evolution.

The main challenge of stacking polarized intensity - a positive
definite quantity - is the treatment of the effects of noise. In the
limit of high signal to noise ratio, a number of corrections have been
proposed to correct for a small bias in the observed polarized
intensity due to noise. The main reason for stacking is to explore the
low signal to noise regime, where the signal at best creates a small
positive bias to the noise. We discuss the effects of noise in
Section~\ref{model-sec} and apply stacking to a simulated survey in
Section~\ref{simulation-sec}. In Section~\ref{NVSS-sec} we apply the
same procedure to NVSS images, and compare our results with the
literature in Section~\ref{discussion-sec}. In the future we will
report on polarization stacking of sub-samples selected using
different surveys.
\\

\section{Stacking polarized intensity}
\label{stack-sec}

\subsection{The data}

Polarization data used here are 1.4 GHz polarization images from the
NVSS that cover 80\% of the sky with a mean sensitivity of 0.29 mJy
beam$^{-1}$ in Stokes $Q$ and $U$ images. The reference catalog for source
positions is the NVSS catalogue, constructed by fitting 2-dimensional
Gaussians to the Stokes I images \citep{condon1998}. The NVSS
catalogue is 90\% complete for sources with $S_{1.4} = 3\ \rm mJy$,
and 50\% complete at $S_{1.4} = 2.5\ \rm mJy$. The rms position error
is $\lesssim 1\arcsec$ for sources brighter than 10 mJy, and better
than $7\arcsec$ (rms) for fainter sources. The original resolution of
the NVSS is $45\arcsec \times 45\arcsec$ (FWHM). However, the NVSS images
were published with discretized pixel values (allowing compression of
the data) that inhibits median stacking to a faction of the
noise. Instead of the published NVSS images, we used the full-band
images made by \citet{taylor2009} for our stacking experiments, with a
resolution of $60\arcsec \times 60\arcsec$ (FWHM). The rms confusion
limit in our total intensity images is $142\ \rm \mu Jy$
\citep{condon1998}, corresponding with $\sim 3\ \rm \mu Jy$ in
polarization for a median percentage polarization of $2\%$.

At low Galactic latitude, the NVSS catalogue is significantly
incomplete in extragalactic sources because of confusion with
small-scale structure in diffuse Galactic emission, and it is
contaminated by fitting of small-scale structure and side lobes
of bright sources. In some regions around the Galactic plane, the NVSS
suffers from bandwidth depolarization \citep{stil2007}. We avoided
significant bandwidth depolarization related to Faraday rotation by
the Galactic foreground by restricting our experiments to sources with
Galactic latitude $|b| > 30\degr$. It is difficult to avoid completely
the small-scale structure in Stokes $Q$ and $U$ in the NVSS
\citep{rudnick2009}, but we found that our conclusions do not change
if the latitude cut-off is increased up to $60\degr$. The software
automatically rejects sources if the local noise level is larger than
a user-specified multiple of the nominal survey noise. We set this
rejection threshold at 3 times the nominal noise value, resulting in a
negligible number of rejections, and found that turning this rejection
off completely had a negligible impact on our results. Visual inspection
of rejected positions confirmed that the rejection was removing poor
areas in the survey, so it was maintained in the final analysis. One
of the great merits of the NVSS is its uniform sensitivity, which is
an important advantage for stacking.

The final input catalogue was divided into narrow bins of 0.05 dex in
total flux density, such that stacking in polarized intensity will
eventually result in a determination of median percentage polarization as a
function of flux density.

\subsection{Construction of stacked images}

\begin{figure*}
\center\resizebox{12cm}{!}{\includegraphics[angle=0]{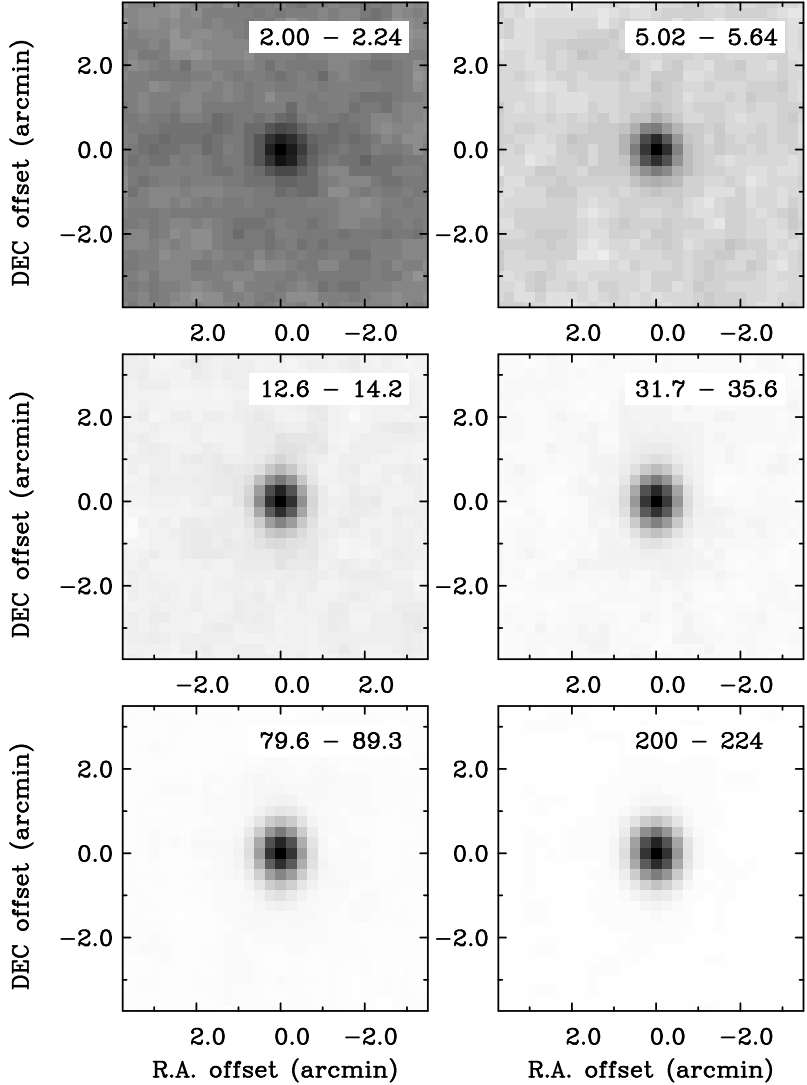}}
\caption{ Sequence of stacked polarized intensity images. The Stokes I
  flux density ranges are indicated in each panel. Grey scales are
  linear from 0.350 mJy beam$^{-1}$ to $p_{\rm med}$ at the central pixel,
  which is 0.386, 0.420, 0.546, 0.906, 1.813, and 4.00 mJy beam$^{-1}$
  respectively. 
\label{stack_images-fig}}
\end{figure*}

The stacked images were made by extracting postage stamp images of $N
\times N$ pixels centered on the catalogued position. We used $N=30$
with $15\arcsec$ pixels in the original images. Aligning the postage
stamps to the nearest pixel creates a significant downward bias by
blurring due to alignment errors in the stacked image. To minimize any
blurring, the postage stamps were aligned by oversampling each by a
factor 8, aligning, and then resampling back to the original pixel
scale of the survey. This reduced any residual alignment errors to
$\sim 1/30$ of the synthesized beam scale, approximately the size of the
rms position errors of the reference catalogue.

\begin{figure}
\resizebox{\columnwidth}{!}{\includegraphics[angle=0]{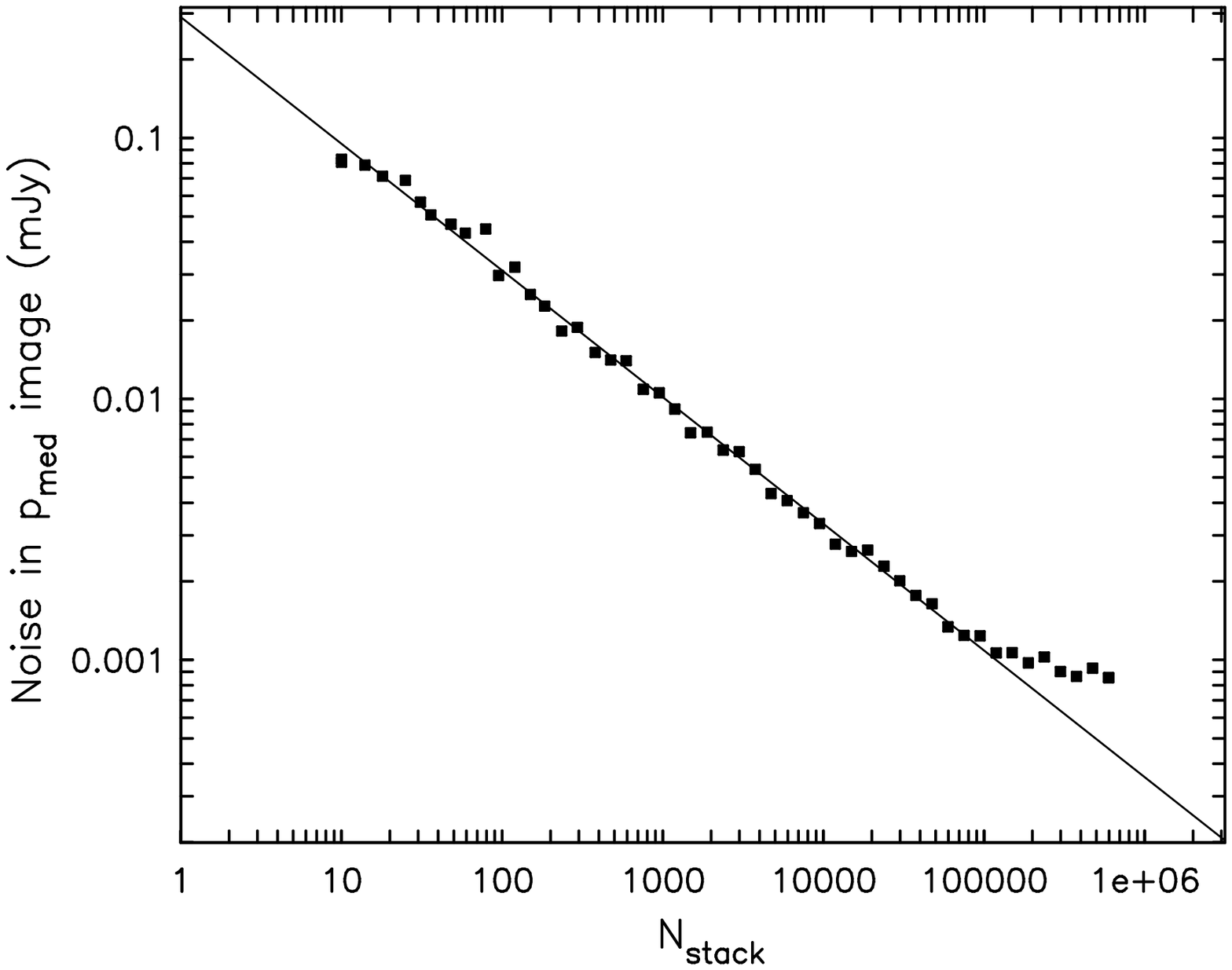}}
\caption{ Reduction of the noise in median stacked images of polarized
  intensity ($p_{\rm med}$) as a function of the number or objects in
  the stack, $N_{\rm stack}$. For this experiment randomly selected
  positions from the area of the NVSS used for stacking polarized
  intensity were used. The line is a fit to the points between $N_{\rm
    stack}=10$ and $N_{\rm stack}=10^5$ with logarithmic slope
  $-0.485$ and intercept $\log[0.291/\rm (mJy beam^{-1})]$. The
  average noise in $Q$ and $U$ images from the NVSS is 0.29 mJy
  beam$^{-1}$. Note the break near $N_{\rm stack} \gtrsim 10^5$. The
  faintest flux density bins in the stacking analysis contains
  $7.7\times 10^4$ sources.
\label{PI_random_positions-fig}}
\end{figure}

The noise in each postage stamp in $Q$ and $U$ was determined by
taking the median of the absolute value of pixels in the border of the
postage stamp, covering approximately 12 independent beams.  This is
more robust than the rms value if another source occurs in the
area. For Gaussian noise with standard deviation $\sigma_{QU}$, the
median of the absolute value is $0.671 \sigma_{QU}$.  Postage stamps
with noise level more than three times the typical noise of 0.29 mJy
beam$^{-1}$ were discarded. The non-Gaussian statistics of the noise
in $Q$ and $U$ were determined empirically from a set of 80
high-latitude mosaics, blanking out pixels with detectable Stokes $I$
emission. We fitted a combination of a Gaussian and exponential wings
to describe the statistics of the noise empirically
\citep{george2012}. The noise distribution was scaled for each target
source to have a median absolute value equal to that measured at the
border of the postage stamp. For each stack we verified that the
off-source stacked polarized intensity is consistent with a set of
Monte-Carlo realizations for zero signal within the statistical
uncertainty. The effectiveness of this approach is best illustrated by
its performance in the stacking simulation presented in
Section~\ref{simulation-sec}, for which we determined the noise
statistics in the same way from the images.

Stacked images in total intensity and polarized intensity were
constructed by taking the median over all remaining postage stamps for
each pixel. When we discuss the median fractional polarization from
stacking, we use the ratio of the median $p_0$ to the median of $I$
for narrow ranges of $I$. Figure~\ref{stack_images-fig} shows a
sequence of stacked polarized intensity images for selected flux
density ranges starting with the faintest flux density, increasing by
a factor 2.5 in each step. Significant polarized emission is visible
in the center of each image, increasing in strength as the median
total flux density increases. The angular diameter of sources in the
stacked Stokes I images is $59\farcs53$ with standard deviation
$0\farcs47$, consistent with the $60\arcsec$ beam size of the images
made by \citet{taylor2009}.

The background in the $p_{\rm med}$ images is positive, representing
the polarization bias in the absence of signal. The median of the Rice
distribution in the absence of signal and noise of 0.29 mJy
beam$^{-1}$ is 0.3415 mJy beam$^{-1}$. The actual background level is
closer to 0.365 mJy beam$^{-1}$, because the noise level in the NVSS Q
and U images is not uniform, and possibly because of non-Gaussian
wings of the noise distribution. We tested improvement of the noise
with sample size $N_{\rm stack}$ by stacking randomly selected
positions.  Figure~\ref{PI_random_positions-fig} shows the rms around
the mean offset level in stacked polarized intensity images at random
positions. For $N_{\rm stack} < 10^5$ the rms fluctuations follow the
relation $0.29/\sqrt{N_{\rm stack}}$. At higher $N_{\rm stack}$, the
noise in the stacked polarized intensity images decreases at a slower
rate. The noise level at the break is close to the expected confusion
noise level in polarization for a median percentage polarization of
$2\%$. While polarized sources at the confusion limit blend together
and partially depolarize, the stacked image remains crowded with
sources with a density of the sky that corresponds with a few beams
per source.  For our stacking result, we have no concern regarding
confusion in polarization because the reference catalogue of NVSS
sources remains well above the confusion limit. The faintest flux
density bin in our analysis contains $7.7\times 10^4$ sources yielding
a sensitivity of 1 $\mu$Jy in the stacked image.

\subsection{Polarization bias correction}
\label{model-sec}

\subsubsection{Bias in stacked polarized intensity}

Noise in polarized intensity biases the observed polarized intensity
$p = \sqrt{Q^2+U^2}$ upward from the actual polarized intensity
$p_0$. If the signal to noise ratio is low, the relation between $p_0$
and the expectation value of $p$, or the median value $p_{\rm med}$,
deviates most strongly from a linear relation. Following
\citet{simmons1985}, the probability density function of the polarized
intensity $p$ for a source with true polarized intensity $p_0$ and the
same but statistically independent Gaussian noise in the Stokes
parameters $Q$ and $U$ is the Rice distribution \citep{rice1945,vinokur1965}
\begin{equation}
F(p\,|\,p_0, \sigma_{QU}) = {p\over \sigma_{QU} } \exp\Bigl[ -{p^2 + p_0^2 \over 2 \sigma_{QU}^2}  \Bigr]J_0\Bigl({i p p_0 \over \sigma_{QU}^2}\Bigr),
\end{equation}
with $J_0(x)$ the zeroth order Bessel function, and $i$ the unit for
imaginary numbers. Consider a number of independent measurements with
the same $p_0$. This is not a realistic assumption in many stacking
experiments, but it provides a helpful analytic expression and may
proof to have a useful application in stacking polarized intensity for
a number of subsequent frequency channels for the same source. The
median $p_{\rm med}$ is defined by the relation
\begin{equation}
\int_0^{p_{\rm med}} F(p\,|\,p_0, \sigma_{QU}) dp = {1 \over 2} \int_0^{\infty} F(p\,|\,p_0, \sigma_{QU}) dp.
\label{median-eq}
\end{equation}

In the limit of low signal to noise ratio in polarization ($p_0
\lesssim \sigma_{QU}$), we approximate the Bessel function by a
polynomial approximation to second order $J_0 (x) \approx 1 - {1\over
  4} x^2$ for $x < 1$ \citep{millane2003}. Substituting this approximation in
Equation~\ref{median-eq} yields an equation for the median in the case
that $p_0 \lesssim \sigma_{QU}$.  Expressed in the normalized
variables $u_0 = p_0/\sigma_{QU}$, $w = p_{\rm med}/\sigma_{QU}$, this
equation is
\begin{equation}
\exp[-w^2/2](1 + {1 \over 2} u_0^2 + {1\over 4} u_0^2 w^2 ) - {1\over 2} - {1\over 4}u_0^2 = 0.
\label{solve_median-eq}
\end{equation}
For $u_0 = 0$ we find the exact solution for the median polarized intensity in the absence of signal, $w_0$,
\begin{equation}
w_0 = \sqrt{2 \ln 2},
\end{equation}
where $\ln$ is the natural logarithm. For $u_0 > 0$, Equation~\ref{solve_median-eq}
gives an approximation for $w$ that deteriorates for higher values of $u_0$. 
Substituting $1/2 = \exp(-w_0^2/2)$ in Equation~\ref{median-eq}, and approximating
\begin{equation}
\exp \Bigl[{w^2-w_0^2 \over 2}  \Bigr] \approx {1 + {w^2 - w_0^2 \over 2}},
\end{equation}
solving for $w$ yields
\begin{equation}
w = w_0 \sqrt{1 + {1\over 2} u_0^2}.
\label{median_est-eq}
\end{equation}

\begin{figure}
\resizebox{\columnwidth}{!}{\includegraphics[angle=0]{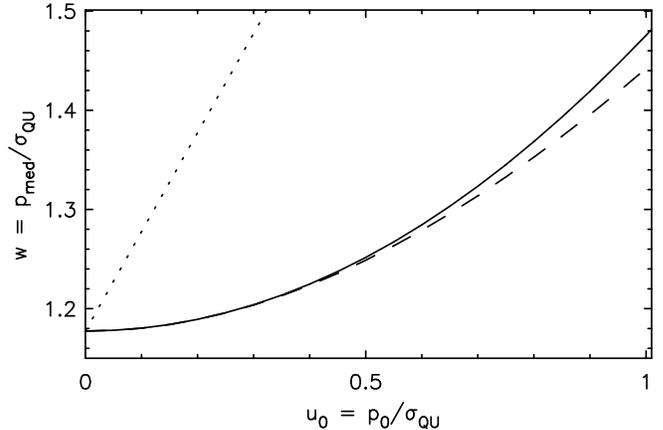}}
\caption{ Dependence of the median polarized intensity on the true
  polarized intensity, for a ficticious sample of sources with the
  {\it same} true polarized intensity $p_0$, and the same noise level
  $\sigma_{QU}$. The solid curve shows the median from numeric
  integration of Equation~\ref{median-eq}. The dashed curve shows the
  approximation in Equation~\ref{median_est-eq}. The difference
  between the two curves does not introduce a significant error when
  stacking modest samples ($N \lesssim 1000$) of $p_0 \lesssim 1
  \sigma_{QU}$ sources. The dotted line represents the
  relation $w = w_0 + u_0$, which is the assumed relation if
  one subtracts the off-source median from the on-source median in an
  attempt to correct for polarization bias.
\label{rice_median-fig}}
\end{figure}

Figure~\ref{rice_median-fig} shows this approximate formula in
comparison with the result of numerically integrating
Equation~\ref{median-eq}. It is to be expected that the approximation
deteriorates for larger $u_0$. The difference remains within the
statistical errors for samples $N \lesssim 1000$ with $p_0 \lesssim
\sigma_{QU}$, for which the noise in the stacked image is more than
$0.03 \sigma_{QU}$. We will see later that a larger source of error is
related to the fact that a real sample of sources contains a
distribution of $p_0$ that is not known a priori. The dotted line in
Figure~\ref{rice_median-fig} is the line $w = w_0 + u_0$. Its
deviation from the curves illustrates that the signal in the stacked
polarized intensity image is far from an addition of the source signal
and the off-source median. Inverting Equation~\ref{median_est-eq}, we
obtain an estimator for the bias correction for the median polarized
intensity, assuming all sources have {\it the same} $u_0 =
p_0/\sigma_{QU} \lesssim 1$,
\begin{equation}
u_0 = 2 \sqrt{{w^2\over 2 \ln 2}-1}.
\label{p0_est-eq}
\end{equation}

The non-linear dependence of $p_{\rm med} = w \sigma_{QU}$ on the true
signal $p_0 = u_0 \sigma_{QU}$ is evident in
Equation~\ref{median_est-eq}. When stacking a set of sources with
different values of $p_0$, sources with higher $p_0$ contribute
more strongly to raising the median of the sample. A significant range
in $p_0$ is to be expected when stacking a sample of sources. Considering
the additional complication that the noise in a survey may not be uniform,
we expect that an analytical solution as derived above leaves
systematic errors that will be larger than the statistical uncertainty
in the median polarized intensity of a large sample of sources.

We apply Monte Carlo simulations to derive a median $p_{0,\rm med}$
from the median of the stack $p_{\rm med}$.  Consider the stack as a
sequence of observed polarized intensities $p_i$, from sources with
true polarized intensity $p_{0,i}$ ($1 \le i \le N_{\rm stack}$). The $p_i$
are realizations of a single experiment of adding noise in Stokes $Q$
and $U$, and then constructing $p_i=\sqrt{Q_i^2+U_i^2}$ for each
source. The local noise level at each source in the stack is recorded,
and the effect of the noise on the $p_i$ can be simulated by a large
number of Monte-Carlo realizations of the stack. The most likely value
for $p_{0,\rm med}$ and its errors are derived from these Monte Carlo
simulations.

A significant complication is that we cannot assume that all sources
in a stack have the same $p_0$. The expectation value or the median of
the observed signals $p_i$ is a non-linear function of the actual
signals $p_{0,i}/\sigma_{QU,i}$ (Figure~\ref{rice_median-fig} and
Equation~\ref{median_est-eq}). The result of this non-linearity is
that sources with higher $p_{0,i}$ in a survey with uniform noise are
more effective raising $p_{\rm med}$ than sources with low
$p_{0,i}$. The result of the stack therefore depends on the shape of
the distribution of $p_{0,i}$ values, not just the median $p_{0,\rm
  med}$.

\begin{figure}
\resizebox{\columnwidth}{!}{\includegraphics[angle=0]{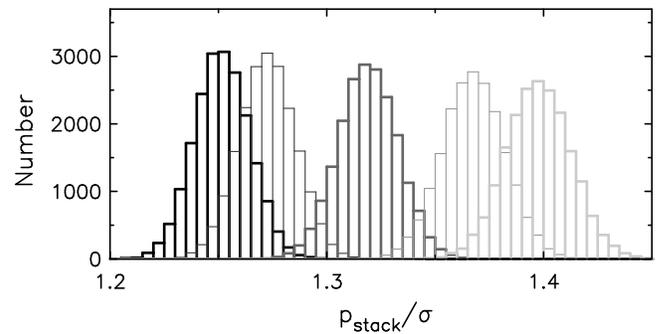}}
\caption{Effect of the shape of the distribution of fractional
  polarization on the outcome of a stack in the presence of
  noise. Shown are the distributions of the outcome of a stack,
  $p_{\rm med}$, in order from left to right, for $p_0$ distributions
  that are a delta function, a uniform distribution, a Gaussian with
  peak at 0, and a Gauss-hermite function with $h_4 = 0.15$ and peak
  at 0, and the $\Pi_0$ distribution from \cite{bg04}.  For each $p_0$
  distribution, the median is $p_{0,\rm med} = \onehalf \sigma_{QU}$, where
  $\sigma_{QU}$ is the noise in Stokes $Q$ and $U$. Equation~\ref{median_est-eq} applies to the delta function distribution and yields $p_{\rm med}/\sigma_{QU}=1.25$
\label{PI_dist_sim-fig}}
\end{figure}

\begin{figure*}
\center\resizebox{8.1cm}{!}{\includegraphics[angle=0]{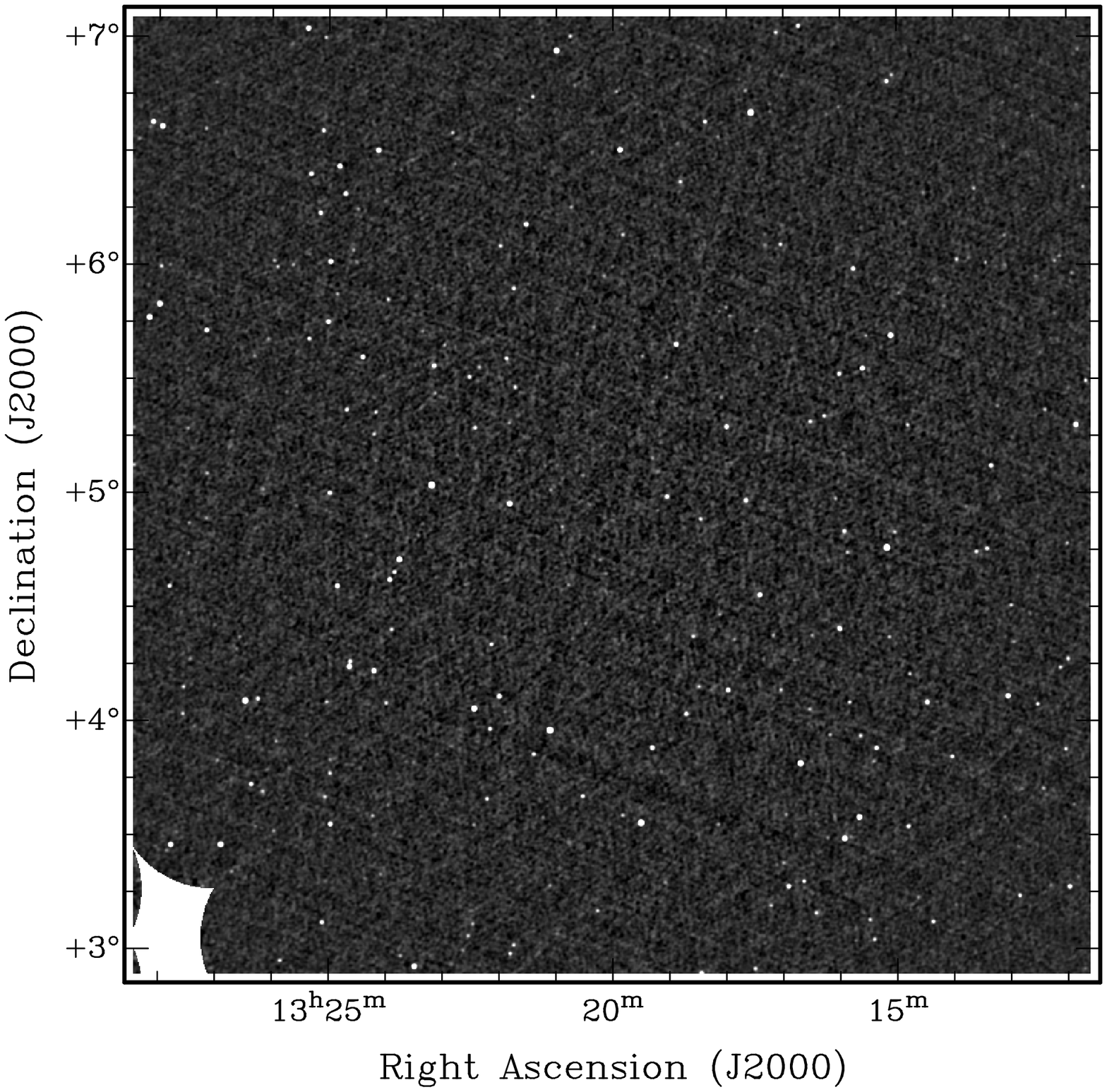}}
\resizebox{8.1cm}{!}{\includegraphics[angle=0]{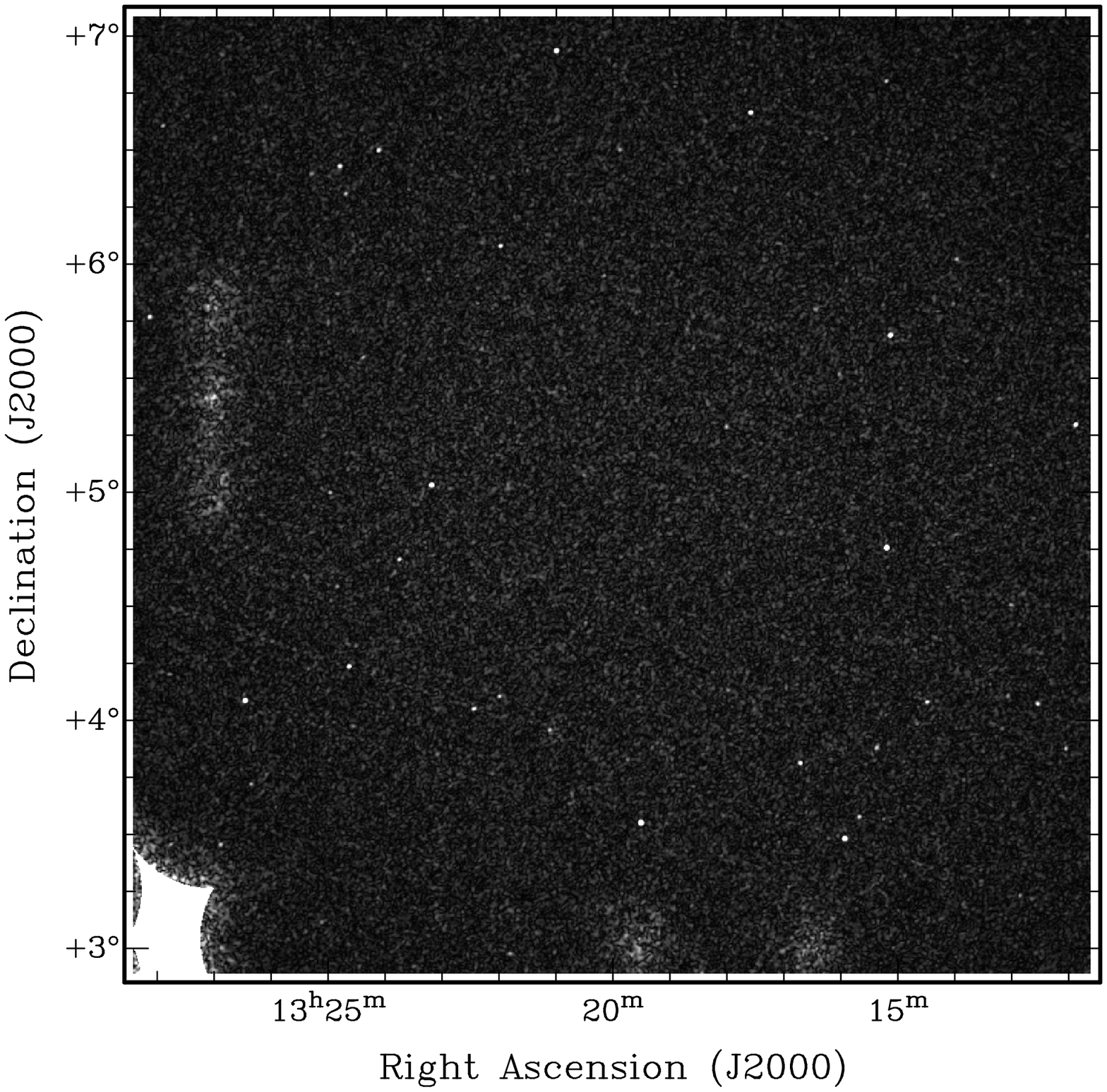}}
\caption{One of 600 simulated $4\degr \times 4\degr$ fields used to
  test polarization stacking with the NVSS. Left: total intensity,
  right: polarized intensity. Some images, such as this one, are
  incomplete and show varying noise level as a result of missing
  fields. Artifacts from sidelobes of bright sources are most apparent
  in total intensity, but exist also in the polarization images. The
  coordinates are artificial and differ from one simulated field to
  another.\label{sim_NVSS-fig}}
\end{figure*}

Figure~\ref{PI_dist_sim-fig} demonstrates the effect of different
$p_0$ distributions with the same median $p_{0,\rm med}$. We simulated
stacking of five samples of 5000 sources and uniform Gaussian noise in
Stokes $Q$ and $U$ with standard deviation $\sigma_{QU}$. We note that
our analysis below takes into account variation of the noise and the
actual noise statistics of the survey that may or may not be Gaussian.
The median polarized intensity of each simulation was the same at
$p_{0,\rm med} = 0.5 \sigma_{QU}$, but the distribution of the
$p_{0,i}$ was different for each sample: all sources the same $p_0$, a
uniform distribution between $p_0 = 0$ and a maximum $p_{0,max}$, a
Gaussian distribution, a Gauss-Hermite function \citep{vdmarel1993}
with extended tail (kurtosis parameter $h_4 = 0.15$, where $h_4 = 0$
for a Gaussian), and the distribution of $\Pi_0$ for bright NVSS
sources derived by \citet{bg04}. The latter is a piecewise fit to the
fractional polarization distribution for NVSS sources brighter than 80
mJy. In addition, a fraction of sources is considered unpolarized. In
this example, that fraction is zero. \citet{bg04} adopted 11.3\%.
Each distribution was stacked in 2000 independent realizations in
order to derive the distribution of the median-stacked polarized
intensity $p_{\rm med}$. The resulting distributions of $p_{\rm med}$
are shown in Figure~\ref{PI_dist_sim-fig}. The total polarization bias
in each stack is the difference between the median of each of the
distributions shown and the intrinsic median $p_{0,\rm med} = 0.5
\sigma_{QU}$.

Although the differences between the distributions in
Figure~\ref{PI_dist_sim-fig} are small compared with the total bias,
they are significant compared with the width of each of the
distributions, which is of the order of $\sigma_{QU}/\sqrt{N}$, with
$N=5000$ in this example. In order to take optimal advantage of the
increased sensitivity in the stacked images, our bias correction must
include the distribution of the $p_0$. This is a higher-order
correction to the derived median $p_{0, med}$ that to some degree
depends on the signal to noise ratio per object. We will include a
constraint to the shape of the $\Pi_0$ distribution in our analysis
below. It is worth noting that for small samples, $N \lesssim 100$,
the differences between the histograms become similar to the
statistical error $\sigma_{QU}/N$. For these small samples, one can
apply Equation~\ref{p0_est-eq} as a bias estimator. The implied delta
function distribution tends to underestimate $p_{\rm med}$ for a given
$p_{0,\rm med}$ compared with distributions of finite width, so this
estimator is biased toward higher $p_{0,\rm med}$ for a given $p_{\rm
  med}$.

\subsubsection{Bias correction of stacked polarized intensity}
\label{bias-sec}

When stacking sources selected in a narrow range of total intensity,
the distribution of $p_0$ is of the same form as the distribution of
fractional polarization $\Pi_0$, which is at best slowly varying with
flux density as the relative numbers of radio source populations
change gradually with flux density.  The $\Pi_0$ distribution derived
for bright sources can be used as an initial estimate for the
$p_{0,i}$ distribution if the sources span a narrow range in total
flux density. Note that this solution works well for samples that
can be binned by their Stokes I flux density.  When stacking targets
selected at a different wavelength, such as a different radio
frequency, X-ray, or optical source catalogue, an a-priori estimate
for the distribution of the $p_{0,i}$ is much more complicated.

In order to proceed, an additional constraint for the {\it
  distribution} of the $p_{0,i}$ is required. When we refer to a fixed
shape of the $\Pi_0$ distribution, we still allow for a scaling in
$\Pi_0$ that changes the median. While this is an approximation,
it allows us to use prior information from brighter sources as a basis
for our analysis. A constraint for the shape of the distribution is
available if the number of sources $N_{\rm stack}$ is sufficiently
large. The distribution of the observed $p_i$ values contains
information about the distribution of the $p_{0,i}$, even though it is
broadened by the effects of noise. Our Monte Carlo simulations
reproduce the effect of the noise in polarized intensity for a
proposed distribution of $p_{0,i}$.  The distribution of the
synthesized polarized intensities including noise can be compared with
the distribution of observed $p_i$. The metric by which we decide
whether the assumed distribution of $p_{0,i}$ is acceptable, is the
ratio of the upper quartile to the lower quartile of the distribution
of the observed $p_i$. This ratio is still fairly robust against
outliers, which is necessary when stacking real data.

For sources selected in a narrow range of total intensity, the shape
of the distribution of observed $p$ approaches the $\Pi_0$
distribution in the limit of high signal to noise ratio, while it
approaches the distribution of the noise in the limit of no polarized
signal. In between these two limiting cases, we find a transition 
between the two distributions that occurs over an order of magnitude
in $p_{\rm med}$ or more. The ratio of the upper quartile to the lower
quartile of the observed $p$ values of the sources in the stack is
sensitive to the shape of this distribution, but not to the median
$p_{\rm med}$ itself. It therefore provides an additional constraint to
the shape of the distribution that is not sensitive to outliers,
provided $N_{\rm stack}$ is sufficiently large. 
 
Polarization bias correction proceeds as follows.  For each source in
the stack, we draw a $\Pi_0$ and polarization angle, and noise
perturbations to simulate observed $Q_i$ and $U_i$ ($i = 1 \ldots
N_{\rm stack}$).  The resulting set of polarized intensities $p_i$ is
stacked, retrieving $p_{\rm med}$ and quartiles. Repeat this $\sim
2000$ times to define the distribution of $p_{\rm med}$ for the input
parameters. This procedure must be repeated for a sequence of assumed
median values $p_{0,\rm med}$ and distributions of $\Pi_0$ until the
quartile ratio matches that of the data in the stack for every flux
density bin.  The input parameters vary little from one flux bin to
another, allowing targeted search ranges for these parameters to be
defined.  Retrieving the $\Pi_0$ distribution from the observed $p_i$
distribution is a deconvolution problem that can in principle be
solved separately, with the result feeding back into the stacking
analysis. This may be necessary if one finds that the $\Pi_0$
distribution derived from bright sources does not reproduce the
quartile ratio derived from the data at lower flux densities.  The
signal to noise ratio for which any particular shape can be
confidently rejected or accepted in this way depends on the size of
the samples. While a complete analysis of all possible shape
distribitions is beyond the scope of this paper, we illustrate this
further in the following sections.

\subsection{A stacking simulation}
\label{simulation-sec}

\begin{figure}
\center\resizebox{\columnwidth}{!}{\includegraphics[angle=0]{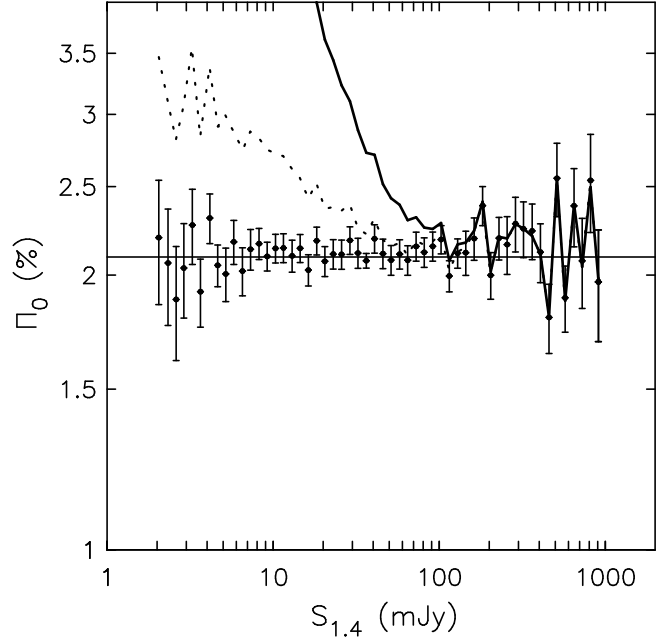}}
\caption{ Fractional polarization as a function of flux density from
  our simulated survey.  The horizontal line marks the median $\Pi_0$
  in the simulations. Error bars are derived from the Monte Carlo
  simulations and represent the 16.5 and 83.5 percentiles of
  distributions similar to those shown in
  Figure~\ref{PI_dist_sim-fig}. The errors are smallest at
  intermediate flux densities. For bright sources, the errors are
  larger because the sample size is smaller, while for faint sources,
  the errors are larger because the signal to noise ratio in $p_{\rm
    med}$ is smaller.  The continuous grey curve represents $p_{\rm
    med}/I$, indicating the magnitude of the polarization bias
  correction for every flux bin. The dotted curve shows results from Monte Carlo
  simulations using a Gaussian $\Pi_0$ distribution, that is rejected
  because it does not fit the quartile ratio in
  Figure~\ref{sim_NVSS_quartile-fig}. 
\label{sim_NVSS_stack-fig}}
\end{figure}

\begin{figure}
\center\resizebox{\columnwidth}{!}{\includegraphics[angle=0]{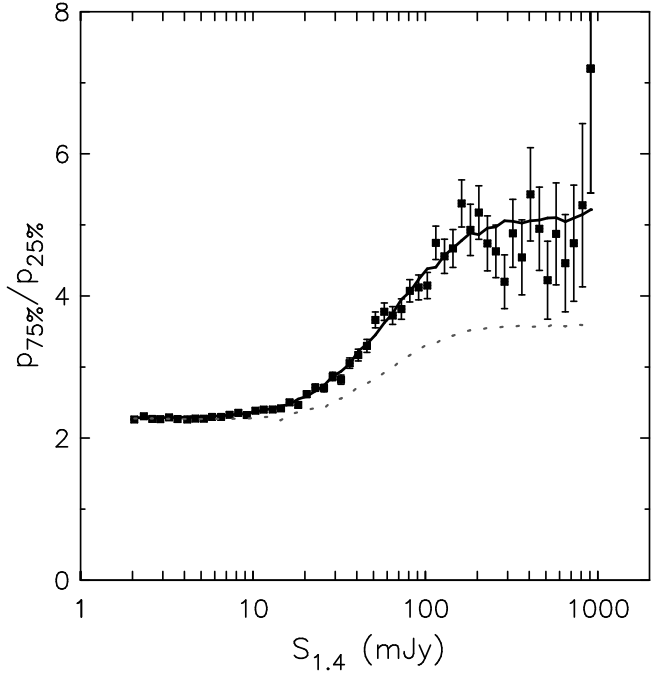}}
\caption{ Distribution of the ratio of upper quartile ($p_{75\%}$) to
  lower quartile ($p_{25\%}$) of the $p_i$ distribution of stacks of
  simulated images as a function of total flux density. The solid
  black curve represents the quartile ratios for the preferred $\Pi_0$
  distribution. The dotted curve shows the quartile ratio for Monte
  Carlo simulations that assume the intrinsic distribution of $\Pi_0$
  is Gaussian (see also dotted curve in
  Figure~\ref{sim_NVSS_quartile-fig}). This Gaussian distribution is
  rejected for sources with $S_{1.4} \gtrsim 10\ \rm mJy$ in this simulation.
\label{sim_NVSS_quartile-fig}}
\end{figure}

In order to test the stacking and bias correction, simulated images
such as the one shown in Figure~\ref{sim_NVSS-fig} were made, that
contain image artifacts and spatial variation of the noise similar to
that encountered in NVSS images. Total flux densities in the range
0.1 mJy to 1 Jy were drawn from a $S^{-2.5}$ power law
distribution. The percentage polarization of each source is drawn from
a distribution fitted to NVSS sources brighter than 80 mJy by
\citet{bg04}. The inserted median $\Pi_{0,\rm med} = 2.1\%$ is independent of
flux density.  Each source is assigned an arbitrary polarization
angle, dividing the polarized signal over Stokes $Q$ and $U$. To simulate
imperfect imaging each source is added to the simulated image in a way
that mimics interferometric imaging with a finite clean limit of $5$
mJy beam$^{-1}$ and a Gaussian restoring beam. Residual side lobes are
appropriately scaled VLA-D antenna patterns centered on the location
of the source.  These side lobes add incoherently, creating a
non-Gaussian noise floor after which Gaussian noise modulated by the
sensitivity pattern of NVSS mosaics is added. We included some weight
patterns of incomplete mosaics that lead to variation in the noise
level around missing fields.  

We stacked polarized intensity following the procedure outlined above.
The results are shown in Figure~\ref{sim_NVSS_stack-fig} and
Figure~\ref{sim_NVSS_quartile-fig}. Each point in
Figure~\ref{sim_NVSS_stack-fig} represents a stack with its own Monte
Carlo analysis. Note that the formal errors in the median fractional
polarization in Figure~\ref{sim_NVSS_stack-fig} are smallest in the
central part of the flux density range, and larger for brighter
sources and for fainter sources. Errors are larger for brighter
sources because the sample sizes are smaller, while for fainter
sources the signal to noise ratio in the stacked image decreases as
the signal gets weaker, despite the larger sample size. The input
median fractional polarization is recovered within the errors over the
entire flux density range shown. To illustrate the magnitude of the
bias correction in the fainter bins, the ratio $p_{\rm med}/I$ is
shown by the gray curve. The dotted curve shows the result of Monte
Carlo simulations that assume a Gaussian distribution of $\Pi_0$,
which is rejected because it does not reproduce the shape of the
distribution of the $p_i$ at intermediate flux densities as explained
below.

Figure~\ref{sim_NVSS_quartile-fig} shows the ratio of the upper to the
lower quartile of the set of $p_i$ for each of the stacks as a
function of flux density (points with error bars), along with a curve
that shows the mean quartile ratio of the 2000 Monte-Carlo simulations
for each stack. The shape of the curve illustrates the transition from
the noise dominated regime to the signal dominated regime. The
zero-signal stack reproduces the off-source values in the stacked
image. Recall that for sources fainter than $\sim 100$ mJy in the
NVSS, more than 50\% of the sources is no longer formally detectable,
leaving the median undetermined unless stacking is attempted. This
analysis cannot provide a unique shape for the distribution of
$\Pi_0$, but it is sufficiently sensitive in the transition region to
differentiate between a Gaussian distribution and the actual
non-Gaussian shape of the distribution of NVSS sources. The dotted
curve in Figure~\ref{sim_NVSS_quartile-fig} shows the quartile ratio
for the Monte Carlo simulations that assumed a Gaussian $\Pi_0$
distribution. A Gaussian $\Pi_0$ distribution is rejected for sources
with $S_{1.4} \gtrsim 10\ \rm mJy$, while the distribution for bright
sources consistently reproduces the quartile ratio over the entire
flux density range. Note that we have to assume that the shape of the
distribution does not change for sources fainter than $\sim$ 10 mJy.
However, we do not expect the shape of the distribution to change
rapidly with flux density.

\subsection{Results from stacking NVSS polarized intensity}
\label{NVSS-sec}
\begin{figure}
\resizebox{\columnwidth}{!}{\includegraphics[angle=0]{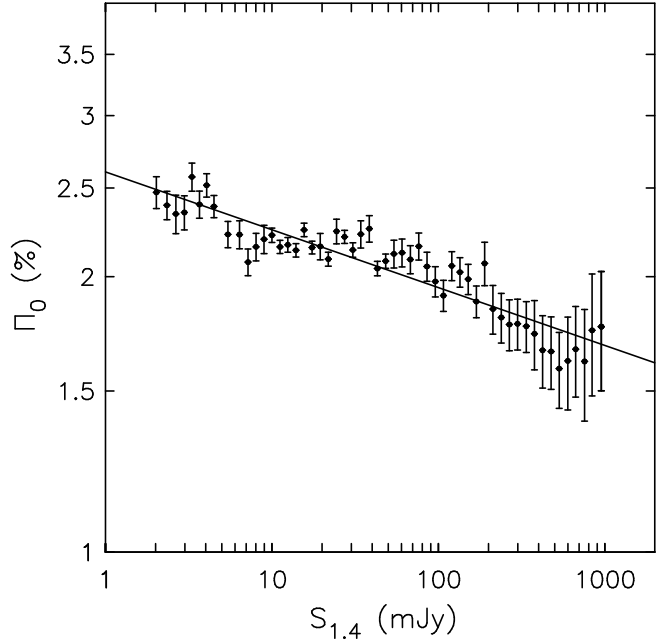}}
\caption{ Fractional polarization as a function of flux density for
  NVSS sources with $|b| > 40\degr$. Each flux density bin
  represents an independent stacking analysis, with no overlap in the
  source lists. The $\Pi_0$ distribution adopted is the \citet{bg04}
  distribution with 6\% unpolarized sources.
\label{NVSS_PIvsI-fig}}
\end{figure}

\begin{figure}
\resizebox{\columnwidth}{!}{\includegraphics[angle=0]{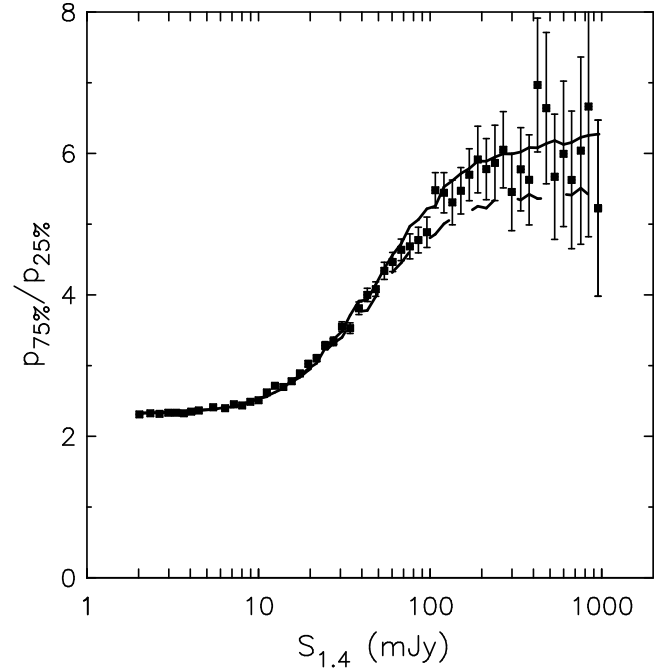}}
\caption{ Distribution of the ratio of upper quartile to lower
  quartile of the $p_i$ distribution of NVSS stacks as a function of
  total flux density. The solid black curve represents a model that
  includes residual instrumental polarization at the 0.3\% (rms)
  level, while the dashed curve represents Monte-Carlo simulations
  without instrumental polarization.
\label{NVSS_quartile-fig}}
\end{figure}

For stacking NVSS sources, we used the distribution fitted by
\citet{bg04}, with 6\% of sources completely unpolarized. The median
fractional polarization of NVSS sources as a function of flux density
is shown in Figure~\ref{NVSS_PIvsI-fig}, and the accompanying quartile
ratio in Figure~\ref{NVSS_quartile-fig}. We find the median fractional
polarization is higher for fainter radio sources. The line in
Figure~\ref{NVSS_PIvsI-fig} represents the fit  
\begin{equation}
\log \Pi_{0,med} = -(0.051 \pm 0.004) \log S_{1.4} + (0.388 \pm 0.007),
\label{pi_vs_I-eq}  
\end{equation}
where $\Pi_{0,med}$ is expressed as percent polarization, and
$S_{1.4}$ is the 1.4 GHz flux density in mJy.  We see small systematic
residuals around this best fit line. While this may indicate real
curvature in the relation, it may also indicate resolution effects for
bright sources that are difficult to quantify. On one side, slightly
resolved sources may be more highly polarized because structure in
polarization angle leads to depolarization in an unresolved source. On
the other side, the stacking centers at the peak of total intensity,
and may miss polarized components that are offset from the peak in
total intensity. A second degree polynomial fit to the data has a
maximum fractional polarization of 2.4\% at flux density $S_{1.4} =
0.4\ \rm mJy$, with considerable uncertainty.  We apply the linear fit
above for our derivation of polarized source counts below, but include
the second order fit in our error analysis.

Figure~\ref{NVSS_quartile-fig} shows the quartile ratio as a function
of flux density for the same stacks. The dashed curve shows the
quartile ratio for Monte Carlo simulations of the stack that included
noise only, while the solid curve also included residual instrumental
polarization, modeled as a random term in $Q$ and $U$ that is Gaussian
with standard deviation $0.3\%$ of Stokes I, similar to the expected
residual leakage in the NVSS. Despite extensive modeling, including
residual leakage was required to fit the quartile ratio over the full
flux density range considered. The NVSS stacking results presented
here account for this leakage term, but we note that the result in
Figure~\ref{NVSS_PIvsI-fig} is not sensitive to this detail because
the median $\Pi_0$ is substantially larger than the residual
instrumental polarization after calibration. In a future paper we will
present a sample of flat spectrum sources stacked in the same way with
median $\Pi_0$ below $1\%$, independent of flux density. In that case,
it is more important to account for residual instrumental
polarization.

\section{Polarized radio source counts}

\begin{deluxetable}{lrr}
\tablecolumns{3}
\tablewidth{0pc} 
\tablecaption{ Polarized source counts for AGN from stacking$^a$ \label{sourcecounts-tab} }
\tablehead{\ \ \ \ \ $p_0$   &\ \ \ \ \ \ \ \ \ \ \ \ \ \ \ \ \ \ \ \ \ $p_0^{5 \over 2}{dN \over dp_0}$\ \ \ &\ \ \ \ \ \ \ \ \ \ \ \ \ \ \ \ \ \ \ \ \   $N(>p_0)$ \\
\ \ (mJy)   & \ \ \ \ \ \ \   $\rm Jy^{3 \over 2}\ sr^{-1}$  &\ \ \ \ \ \ \ \ \ \ \ \ \ \ \ \ \ \ \ \ \ $\rm sr^{-1}$ \\}
\startdata
     0.0050  &    0.0103  &   $1.02 \times 10^6$  \\
     0.0083  &    0.0133  &   $6.46 \times 10^5$  \\
     0.0138  &    0.0175  &   $4.21 \times 10^5$  \\
     0.0229  &    0.0241  &   $2.80 \times 10^5$  \\
     0.0380  &    0.0340  &   $1.88 \times 10^5$  \\
     0.0631  &    0.0489  &   $1.27 \times 10^5$  \\
     0.1047  &    0.0706  &   $8.60 \times 10^4$  \\
     0.1738  &    0.1027  &   $5.80 \times 10^4$  \\
     0.2884  &    0.1503  &   $3.90 \times 10^4$  \\
     0.4786  &    0.2217  &   $2.59 \times 10^4$  \\
     0.7943  &    0.3295  &   $1.68 \times 10^4$  \\
     1.318  &    0.4828  &   $1.05 \times 10^4$  \\
     2.188  &    0.6814  &   $6.31 \times 10^3$  \\
     3.630  &    0.9115  &   $3.58 \times 10^3$  \\
     6.026  &    1.135  &   $1.92 \times 10^3$  \\
    10.00  &    1.320  &   $9.79 \times 10^2$  \\
    16.60  &    1.454  &   $4.80 \times 10^2$  \\
    27.54  &    1.572  &   $2.27 \times 10^2$  \\
    45.71  &    1.690  &   $9.85 \times 10^1$  \\
    75.86  &    1.626  &   $3.62 \times 10^1$  \\
   125.89  &    1.235  &   $1.05 \times 10^1$  \\
   208.93  &    0.724  &   $2.29 \times 10^0$\  \\
   346.74  &    0.283  &   $0.261 \times 10^{0}$  \\
\enddata
\tablenotetext{a}{Source counts for $p_0\gtrsim 50\ \rm \mu Jy$ are directly constrained by the results shown in Figure~\ref{NVSS_PIvsI-fig}, while source counts for $p_0 \lesssim 50\ \rm \mu Jy$ rely on an extrapolation of the relation shown in Figure~\ref{NVSS_PIvsI-fig} (see text for details).}
\label{sourcecounts-tab}
\end{deluxetable}

Equation~\ref{pi_vs_I-eq} can be combined with radio source counts in
total intensity to predict polarized radio source counts in a similar
fashion as \citet{bg04}. Stacking allows us to probe the fractional
polarization of radio sources more than an order of magnitude deeper
than previous work based on the NVSS, probing the flux density range
where most AGN in the radio source counts have radio luminosity
below the traditional FRI/FRII luminosity boundary
\citep[e.g.][]{wilman2008}.

We combine the stacking results with total intensity source counts
from the SKADS S3 simulation of \citet{wilman2008}. These model source
counts fit observed radio source counts well, except maybe at the very
bright end which is less important for our purpose. The models also
provide a physically motivated extrapolation to lower flux
densities. We used all sources of type AGN, adding the fluxes of
multiple components where appropriate. For the flux range that we
consider here, these Stokes I source counts consist of FR I and FR II
class sources with a minor contribution of radio-quiet QSOs at the
faint end of the flux density range considered here.

Equation~\ref{pi_vs_I-eq} is a fit to data for sources with flux
density $S_{1.4} \ge 2\ \rm mJy$.  A median percentage polarization of
$\Pi_{0,med} = 2.5\%$ corresponds to a median polarized flux density
of $50\ \rm \mu Jy$. For polarized flux densities $p_0 \gtrsim 50\ \rm
\mu Jy$, the polarized source counts are therefore directly
constrained by stacking.  Below this flux density, our tabulated
polarized source counts represent an extrapolation
of Equation~\ref{pi_vs_I-eq}. We do not expect a sudden
change in polarization properties of radio sources at any flux density, 
because radio sources at any flux density include sources
with a wide range in luminosity and red shift. This "convolution with
the universe'' justifies extrapolation by a factor $\sim10$ in flux
density. The polarized source counts we derive in this
flux density range will not be seriously affected by an emerging
polarized source population at the faint end. 

Table~\ref{sourcecounts-tab} lists Euclidean-normalized differential
source counts and cumulative source counts derived by convolving the
AGN total-intensity source counts from \citet{wilman2008} with the
distribution of $\Pi_0$ derived from stacking. These models provide us
with an AGN-only version of radio source counts that matches the
population of radio sources stacked. For the purpose of this paper, it
is sufficient that these models fit observed source counts in the flux
density range of interest, while the cosmological details may only
affect the results insofar they rely on separation of different
populations of radio sources.

The uncertainty in the polarized source counts is a result of
uncertainty in the AGN source counts from the \citet{wilman2008}
simulation, and uncertainties in the results from stacking.  In the
flux density range (2 mJy - 30 mJy), the median fractional
polarization is directly constrained by stacking, and the AGN
component of the source counts in the SKADS S3 simulations is
well-constrained by observations. According to the models, these AGN
are dominated by radio sources with luminosity consistent with type FR
I.  Below $\sim 1$ mJy, the total source counts are increasingly
dominated by star forming galaxies, and observed source counts from
small deep fields show considerable spread. While a significant
fraction of radio sources with 1.4 GHz flux density in the range 0.1 -
1 mJy is AGN related \citep{gruppioni1999}, the AGN fraction below 1
mJy becomes gradually more uncertain, introducing uncertainty that
grows with the degree of extrapolation to lower flux densities. The
radio-quiet QSO population in the models represents objects in the
low-luminosity end of the radio luminosity function of AGN that are
not well represented in the flux density range of NVSS sources.  Within
the context of the \citet{wilman2008} models, it appears reasonable to
extrapolate the stacking results over about an of magnitude in flux
density below the faintest flux density bin that was actually
stacked. We therefore apply $S_{1.4} \gtrsim 0.2\ \rm mJy$, or $p_0
\gtrsim 5\ \rm \mu Jy$ as the limits for the polarized source counts,
while noting that the stacking experiment constrains these counts
directly for $p_0 \gtrsim 50\ \rm \mu Jy$. 

Formal errors in the fit given in Equation~\ref{pi_vs_I-eq} result in
an uncertainty on the level of a few percent in the cumulative source
counts at $5\ \rm \mu Jy$ in Table~\ref{sourcecounts-tab}. We estimate
the actual errors in the normalization of the fit to be closer to 5\%
- 10\% considering variation associated with the uncertainty in the
distribution of $\Pi_0$. If we apply a second-order polynomial fit,
the cumulative polarized source counts for $p_0 > 5\ \rm \mu Jy$ are
10\% lower than listed in Table~\ref{sourcecounts-tab}. For comparison,
the cumulative counts at 5 $\mu$Jy would be 30\% lower than listed in
Table~\ref{sourcecounts-tab} if the median fractional polarization is
held constant at 2\%.

\section{Discussion} 
\label{discussion-sec}

\begin{figure*}
\center\resizebox{13.5cm}{!}{\includegraphics[angle=0]{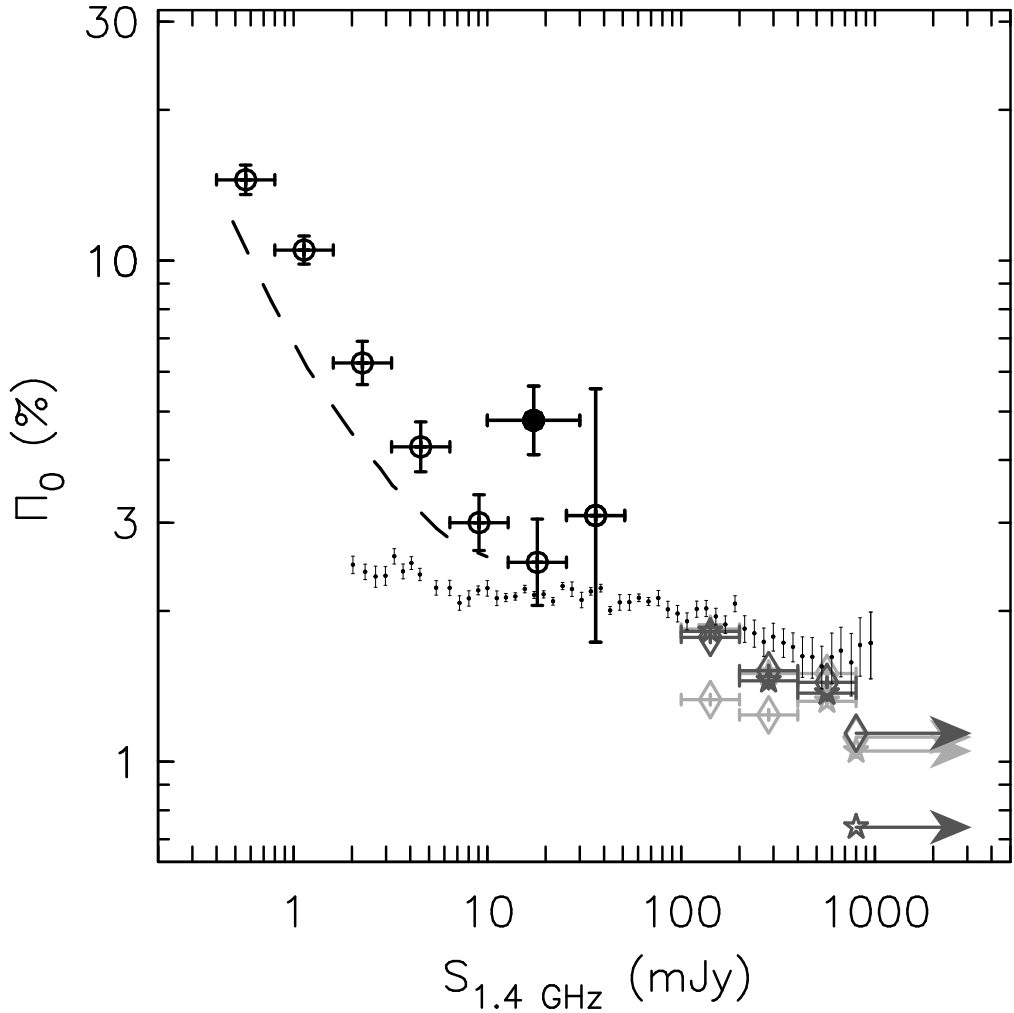}}
\caption{Polarization of radio sources at 1.4 GHz from various sources
  in the literature. The sequence of small points with error bars is a
  reproduction of the result shown in
  Figure~\ref{NVSS_PIvsI-fig}. Stars represent median fractional
  polarization by \citet{mesa2002} and diamonds by \citet{tucci2004},
  both derived from the NVSS (dark grey for steep spectrum, light grey
  for flat spectrum sources). Open circles show median fractional
  polarization from \citet{subrahmanyan2010}. The filled circle shows
  the median fractional polarization from \citet{taylor2007}. The
  dashed curve shows the result of a simulation of the analysis by
  \citet{subrahmanyan2010} for a population of sources with median
  $\Pi_{0,\rm med} = 2.44\%$ as explained in the text.
\label{PI_flux_pub-fig}}
\end{figure*}

Figure~\ref{PI_flux_pub-fig} compares published median $\Pi_0$ as a
function of flux density with the results of this paper. The
main source of data for $S_{1.4} \gtrsim 100\ \rm mJy$ is the NVSS
\citep{condon1998}, with sample sizes $\gtrsim 10^3$ for individual
flux ranges. Data for fainter sources come from various deep
observations of small fields that detected in order of $10^2$ sources
in polarization.

The data in Figure~\ref{PI_flux_pub-fig} show a tendency for higher
fractional polarization in faint sources. Closer inspection reveals
some notable differences between the published values. At flux
densities higher than 100 mJy, the difference in fractional
polarization between flat spectrum sources and steep spectrum sources
noted by \citet{mesa2002} and \citet{tucci2004} is clearly visible. It
is not surprising that our results agree with their steep spectrum
sources, since a large majority of radio sources has a steep spectrum.
A paper presenting stacking polarized intensity for samples selected
by spectral index is in preparation.  The stacking results presented
here are slightly above the medians for steep spectrum sources
reported by these authors, with medians that are $\sim 0.1\%$ higher,
approximately the formal error per data point in the stacking
experiment. The results of \citet{mesa2002} and \citet{tucci2004}
differ mutually by a similar amount, with slightly higher values
reported in \citet{tucci2004}. The NVSS catalogue is strictly speaking
a catalogue of source components, and the differences between
\citet{mesa2002} and \citet{tucci2004} can be ascribed to differences
in source selection \citep[see][for a discussion]{tucci2004}. The
latitude cut-off $|b| > 2\degr$ appears too close to the Galactic
plane since the NVSS has been shown to suffer from localized bandwidth
depolarization up to latitudes $\sim 20\degr$
\citep{stil2007}. Experiments with different latitude cut-offs on the
entire NVSS catalogue suggest that the median fractional polarization
varies on the level of $\sim 0.1 \%$ depending on the choice of the
latitude cut-off. \citet{mesa2002} also found a median fractional
polarization of 2.2\% for all NVSS sources brighter than 80 mJy, which
is in good agreement with the results from stacking. While a small
systematic difference exists, we recall that the stacking result is
based on NVSS images at $60\arcsec$ resolution made by
\citet{taylor2009}, and that the latitude cut-off and the correction
for polarization bias are different. 

The differences with the literature are larger for sources fainter
than $100\ \rm mJy$ at 1.4 GHz. The surveys below $100$ mJy are
different in observational setup, and the results have been derived
using different methods. While it is possible that cosmic variance may
cause differences between the results of small fields, we suspect that
differences in survey parameters (mean signal to noise ratio at a
given polarized flux density, resolution) and details of source
detection and treatment of polarization bias contribute considerably
to the differences between surveys.

The median $\Pi_0 = 4.8 \pm 0.7$ reported by \citet{taylor2007} is the
highest value between 10 mJy and 100 mJy, approximately $5\sigma$
higher than the results of \citet{subrahmanyan2010} and the values
from the present stacking analysis in the same flux density range. The
survey areas, angular resolution, and sensitivity of the surveys of
\citet{taylor2007} and \citet{subrahmanyan2010} are comparable.
\citet{macquart2012} found a mean fractional polarization of 3.7\% for
sources below 20 mJy in the Centaurus field, not shown alongside the
medians in Figure~\ref{PI_flux_pub-fig}. These authors also list 3\%
polarization for sources brighter than 200 mJy, which is substantially
higher than the median of NVSS sources, because it represents a mean,
not a median. If the ratio of mean to median for bright sources can be
applied to faint sources, the median for faint sources in the
Centaurus field would be $\sim 2.5\%$, in the range of $\Pi_{0, \rm
  med}$ from the ATLBS survey and the present analysis around $S_{1.4}
= 10\ \rm mJy$. \citet{macquart2012} fitted a broken power law to the
observed distribution of polarized intensities without the need of a
bias correction. Although their data can be fitted with a significant
change in slope of the polarized source counts, the authors summarized
their result as no turnover above 0.2 mJy.

Before we discuss the higher value from \citet{taylor2007} in this
context, we examine the data for sources fainter than 10
mJy. Below 10 mJy, the medians reported by \citet{subrahmanyan2010}
show a sharp rise toward lower flux density, while the stacking
experiment presented here shows a much more gentle increase. 

The sharp rise in median fractional polarization found by
\citet{subrahmanyan2010} is a result of their polarization bias
correction. \citet{subrahmanyan2010} applied a polarization bias
correction of the form $\hat{p}_0=\sqrt{p^2 - f_p \sigma_{QU}^2}$. The
factor $f_p$ corrects for the size of the aperture in which $Q$ and
$U$ were integrated, and it accounts for correlation of pixel values
on small angular scales arising from convolution with the synthesized
beam.  We illustrate the problem with a simulation, represented by the
dashed curve in Figure~\ref{PI_flux_pub-fig}. Sources were drawn from
the source counts curve of \citet{hopkins2003}, and assigned a $\Pi_0$
drawn from the distribution used in our stacking analysis with
$\Pi_{0,\rm med} = 2.44\%$ (Equation~\ref{pi_vs_I-eq} for $S_{1.4} =
1\ \rm mJy$), independent of flux density. For a random position
angle, Gaussian noise with standard deviation $\sigma_{QU} = 0.085$
mJy beam$^{-1}$ was added to the simulated Stokes $Q$ and $U$. The
$f_p$ factor is difficult to model. It is equivalent to say that the
noise in the integrated $Q$ and $U$ flux density is a factor
$\sqrt{f_p}$ higher than for single-pixel values, and use this higher
noise value in the conventional bias correction $\hat{p}_0=\sqrt{p^2 -
  \sigma_{QU}^2}$. By using only the central noise value $0.085$ mJy
beam$^{-1}$, we implicitly assume $f_p = 1$ for every source and also
a constant noise level in the ATLBS mosaic. This means that the
simulation adds less noise than it should for some sources. As a
result, our simulation does not include some of the more biased
sources in the sample.

The standard polarization bias correction is applied to each
source. If $p< \sigma_{QU}$ the bias-corrected polarized flux density
is assigned the value zero. The resulting median fractional
polarization as a function of flux density is represented by the
dashed curve in Figure~\ref{PI_flux_pub-fig}, which resembles the
steep rise found by \citet{subrahmanyan2010}, even though the true
median $\Pi_{0,\rm med}$ is constant at 2.44\%. The simulation results
remain below the data of \citet{subrahmanyan2010}, but this is at
least in part because the actual noise for some sources must be higher
than the minimum noise used in the simulation. The dashed curve in
Figure~\ref{PI_flux_pub-fig} is not very sensitive to the assumed
slope of the Stokes $I$ source counts on the faint end. Repeating the
simulation with a power law with slope $-1.7$ to $-1.8$ fitted to the
actual counts listed by \citet{subrahmanyan2010} gave the same
results. Setting the slope of the Stokes $I$ source counts to zero
results in a slightly lower curve, approximately $\Pi_{0,\rm med} -
1\%$, indicating a minor contribution of the Eddington bias
\citep{eddington1913,teerikorpi2004}.

We conclude that the main reason for the high fractional polarization
reported by \citet{subrahmanyan2010} is the application of an
incorrect bias correction for sources with a low signal to noise
ratio. Their estimator for $p_0$ has a bimodal distribution for faint
sources, with a peak at zero arising from sources with $p <
\sqrt{f_p}\sigma_{QU}$, and a broader peak around $\sigma_{QU}$. For
faint sources, the relative error in the estimator of $p_0$ is large
and asymmetric \citep[e.g.][]{vaillancourt2006}.

The forward fitting of polarized source counts by \citet{macquart2012}
avoids a correction for polarization bias and naturally accounts for
Eddington bias. These authors warned against considering only the mean
bias correction for analysis.  \citet{macquart2012} fitted a double
power law model of the polarized source counts to the observed
distribution of $p_i$. The model source counts are formulated in terms
of $p_0/\sigma_{QU}$. Interpretation of the result is complicated for
a survey with non-uniform noise. The percentage polarization is
derived by inverting the polarized source counts to the
total-intensity source counts. This requires an assumption of the
shape of the $\Pi_0$ distribution. Our procedure for stacking
polarized intensity includes variation of the noise, and constrains
the distribution of $\Pi_0$ to the extent
possible. \citet{macquart2012} work with results derived from Faraday
Synthesis \citep{brentjens2005}, where the present analysis is
strictly speaking applicable to narrow-band observations. Specific
applications of stacking to wide-band polarization surveys are
discussed briefly in Section~\ref{wideband-sec}.

The Monte-Carlo simulations in \citet{taylor2007} included varying
noise across the mosaic and the slope of the radio source counts in
order to correct for these effects. The cause of the discrepancy
between the present result and the higher median from
\citet{taylor2007} is therefore not immediately clear. The median
fractional polarization $\Pi_{0,\rm med} = 4.8\% \pm 0.7\%$ derived by
these authors, compared with the $10\%$ polarization of {\it detected}
sources with flux density near 10 mJy, indicates that the detected
sources are well in the tail of the $\Pi_0$ distribution, at 2 to 4
times the median. The median fractional polarization in this case may
depend on the adopted slope of the source counts because both can
affect the number of faint sources with polarized flux density higher
than the median. Experiments at the time did not suggest a strong
dependance of the result on the adopted shape of the source counts or
the detection threshold. A more likely potential cause for a higher
median resulting from the analysis of \citet{taylor2007} is a more
implicit effect related to the Eddington bias mentioned above.
Although the Monte Carlo simulations in \citet{taylor2007} were
designed to account for such bias, the likelihood function in their
Equation 6 contains only the error in $p$ associated with the noise in
the image at the position of the source. While the simulated
distributions in $I$ and $p$ did include the effect of noise, the
detection threshold and implicitly the Eddington bias through the use
of the source counts curve in the simulations, sources with high $p$
but low $p/\sigma_{QU}$ due to their position in the mosaic may have
received disproportionate weight in the likelihood function, because
the error bars derived from the noise in the images would not
accurately reflect the width of the probability distribution of $p$
for those sources. This may have lead the maximum likelihood fit to a
higher median.  A re-analysis of the more sensitive ELAIS N1 data from
\citet{grant2010} is required to verify this suggestion.

With stacking we measured the median fractional polarization over
nearly 3 orders of magnitude in flux density. The results are mostly
consistent with previous studies of the NVSS above 80 mJy, and the
trend for higher fractional polarization we find down to $S_{1.4} =
2\ \rm mJy$ appears to be an extension of the trend seen for brighter
sources by \citet{mesa2002} and \citet{tucci2004}. The increase in
fractional polarization is so gradual that it becomes comparable to
the statistical errors for a modest sample size of order 100 per flux
density bin, that may be obtained from a deep integration of a small
area of the sky.  While we find consistency with deep-field results in
their high signal to noise ratio limit, the median fractional
polarization derived here is on the low side of the range of published
values for sources fainter than 100 mJy.  This illustrates how
stacking can supplement observations of small areas with high
sensitivity. 

Establishing the trend in polarization with flux density of AGN
provides an extra constraint that allows us to include magnetic field
properties in models of the cosmic evolution of radio sources through
radio source counts. The cause of the gradual change in $\Pi_{0, \rm
  med}$ with flux density is not known. We speculate that this gradual
change is related to a gradual shift in the radio source population,
for example between steep and flat spectrum sources.  Stacking as a
function of spectral index will be the subject of a subsequent paper.

\subsection{Density of the RM grid}

A key parameter for the scientific impact of all-sky polarization
surveys is the density on the sky of lines of sight with a measured
rotation measure (the RM grid). The polarized source counts presented
here are on the low side of the range considered by
\citet{stepanov2007}, but they agree well with a recent determination
by \citet{rudnick2013}. \citet{macquart2012} sound a slope $-2.17$ for
the differential polarized source counts by fitting a power law count
to polarized intensity distribution of sources not formally detected,
comparable to the $-2.2$ slope they found for detected sources in the
range $0.6 < p < 6\ \rm mJy$. The slope of the differential source
counts in Table~\ref{sourcecounts-tab} is $-1.87$ between $5\ \rm \mu
Jy$ and $100\ \rm \mu Jy$, and $-1.84$ between $0.5\ \rm mJy$ and
$5\ \rm mJy$. The effect of the increase in fractional polarization is
in part compensated by a gradual decrease in the slope of the total
intensity source counts of the same radio sources.

The Polarization Sky Survey of the Universe's Magnetism
\citep[POSSUM,][]{gaensler2010} is designed with a target sensitivity
of $10\ \rm \mu Jy$. Rotation measures can be extracted for sources
with signal to noise ratio in polarization $\gtrsim 10\sigma_{QU}$ or
$p_0 \gtrsim 100\ \rm \mu Jy$. The results in
Table~\ref{sourcecounts-tab} suggest POSSUM will detect approximately
26 polarized sources per square degree extragalactic sky, and fewer
close to the Galactic plane where confusion with the diffuse Galactic
foreground will likely reduce completeness.  If each detected
polarized source above the threshold yields a useful RM measurement,
the mean angular distance between RMs will be about $12\arcmin$. The
RM catalogue from \citet{taylor2009} contains approximately one source
per square degree.

\citet{bg04} assumed in their analysis that only 50\% of polarized
sources provide useful RMs. When comparing the present results with
their extrapolated source counts this factor 2 must be taken into
account. The present polarized source counts at $5\ \rm \mu Jy$ are
then a factor 2 lower than those of \cite{bg04}, even though we find a
gradual increase in fractional polarization for fainter sources.  The
difference is in the adopted total-intensity source
counts. \citet{bg04} used the total source counts from
\citet{hopkins2003} that include a change of slope below 1 mJy that is
generally attributed to star forming galaxies. The polarized source
counts presented here are AGN-only counts. Spiral galaxies may
contribute to polarized source counts at 1.4 GHz, although
depolarization is strong at 1.4 GHz
\citep{stil2009,sun2012}. Polarized source counts models by
\citet{stil2009a} show a significant contribution by spiral galaxies
to the total polarized source counts below $\sim 20\ \rm \mu Jy$, but
these models did not yet take into account an anti-correlation between
fractional polarization and luminosity reported by \citet{stil2009}. 

In summary, the lower fractional polarization we find in this paper
lead to a downward revision of the expected density of the RM grid
from sky surveys with $\mu$Jy sensitivity by a factor of a few. A
recent paper on polarization in the GOODS-N field by
\citet{rudnick2013} comes to a similar conclusion. There is some
latitude in the assumption that only 50\% of sources may yield useful
RMs. For faint sources, statistical errors will dominate over
complexity of the sources. What constitutes a useful RM may differ
between applications.

\subsection{Channel stacking of polarized intensity}
\label{wideband-sec}

Future surveys of polarized radio emission will yield spectral image
cubes of the Stokes parameters that provide the sensitivity of broad
bandwidth but avoid bandpass depolarization. Rotation Measure
synthesis \citep{brentjens2005} is the preferred method to derive
polarization properties of radio sources from these
spectropolarimetric surveys. The detection threshold for Faraday synthesis
for a predetermined acceptable false detection rate is higher because
of the unknown rotation measure of a source
\citep{george2012,macquart2012}. The detection threshold depends on
the Faraday depth range that is searched.

The $\sqrt{N}$ improvement of the noise level we found in the NVSS up
to sample size $N \sim 10^5$ in principle provides sufficient
sensitivity to detect polarization for samples selected at other
wavelengths that remain below the detection threshold in total flux
density. The number of targets is typically much larger than the
number of frequency channels, which will be of order $10^3$ for
surveys with bandwidth of $\sim$1 GHz and frequency channels of
$\sim$1 MHz. The expected sensitivity for stacking a sample of $N \sim
10^5$ in a single frequency channel of a wide-band polarization survey
is therefore an order of magnitude larger than the expected
sensitivity for a single source employing the full bandwidth. The
advantage of high sensitivity in a narrow frequency range is that it
allows investigation of the fractional polarization as a function of
redshift in the rest frame of the target sample, using the bandwidth
of the survey to trace the same emitting wavelength over a signifcant
range in cosmological redshift. This becomes significant for surveys
with the SKA, that will provide large samples of sources at a wide
range of redshifts.  Stacking polarized intensity can thus supplement
Faraday synthesis in investigations of the polarization of radio sources
over cosmic time, as it capitalizes on a unique slice in
multi-dimensional data space. The stacking experiment itself can
address some aspects of Faraday rotation in the source if one stacks
the sample not just in individual channels, but also in channel
averages of polarized intensity, taking both the vector sum and the
scalar sum of polarized intensity over $N_{\rm chan} = 2, 3, 4 \ldots$
adjacent channels across the band. 

Stacking (averaging) polarized intensity per channel in principle can
detect polarization of fainter sources than rotation measure synthesis
with a high significance level.  A difference with the stacking
discussed in this paper is that one can use the mean polarized
intensity in place of the median intensity, because extreme data
values are not expected. The central limit theorem applies and
provides an analytic expression for the distribution of the outcome of
the stack, provided that polarized intensity, or more precisely the
signal to noise ratio in polarized intensity, is trusted to be
constant across the frequency band. If the median is used, the bias
correction in Equation~\ref{p0_est-eq} can be applied to this problem.
If the polarized intensity is not constant, the non-linear relation
between $p_0$ and $p$ introduces similar challenges as described in
this paper for stacking polarized intensity from different
sources. Polarized intensity is likely to change between channels in
surveys with a large bandwidth because its frequency dependence is the
same as total intensity if the fractional polarization is constant,
and because the system temperature may not be constant over the
observed frequency range. If the polarized emission in a source is the
composite of two or more regions subject to different amounts of
Faraday rotation, the vector sum of different RM components can also
lead to complicated variation of polarized intensity with frequency.

Spectral index effects can be removed by dividing polarized intensity
by total intensity.  The statistics of the ratio is not strictly
Ricean because $Q/I$ and $U/I$ will not have Gaussian statistics, even
if the noise in $I$, $Q$, and $U$ is Gaussian. If the signal to noise
ratio in $I$ in a single frequency channel is high, Ricean statistics
may apply, and a correction can be made for polarization bias similar
to that described in Section~\ref{bias-sec} with Monte-Carlo
simulations, or based on the central limit theorem.  If significant
variation in $p/I$ exists, it will be difficult to extract even a mean
or median value because the distribution of intrinsic values must be
taken into account. Splitting up the frequency band into a small
number of sub-bands and stacking each sub-band individually can reduce
this problem, or make it apparent that a more detailed analysis is
required. If the noise is not uniform across the frequency band, one
also needs to model the range of signal to noise ratios in the data.
Failing to do so may lead to systematic errors as a function of signal
to noise ratio.

\section{Conclusions}

Stacking polarized intensity allows us to investigate the faint
polarized signal of radio sources, using large samples covered by
wide-area radio surveys. This technique is already useful at higher
flux densities where a significant fraction of sources is detectable
in polarization. As the sample covers a large area of the sky, results
from stacking are not sensitive to cosmic variance. It opens the
possibility to investigate the polarization of sources with a low
density on the sky, for which narrow deep fields do not provide a
large enough sample. 

We present a procedure for stacking polarized intensity that uses the
shape of the distribution of data values going into the stack as an
additional constraint to solve for the unknown intrinsic distribution
of polarized intensity of the sample.  

We find that the median fractional polarization of sources detected by
the NVSS from stacking polarized intensity is higher for fainter
sources, but the degree of polarization of sources in the flux density
range 2 to 20 mJy remains below 2.5\%, which is significantly smaller
than claimed by previous work.  Polarized radio source counts for
radio sources powered by an active galactic nucleus are derived from
stacking by convolving the $\Pi_0$ distribution with total-intensity
counts modeled by \citet{wilman2008}. These new source counts for $p_0
> 5\ \rm \mu Jy$ are on the low side of the spectrum of predictions
made for the design of future polarization surveys with the SKA and
its path finders, ASKAP and MeerKAT.

\section*{Acknowledgments}
This research has been made possible by a Discovery Grant from the
Natural Sciences and Engineering council of Canada to Jeroen Stil.
JMS thanks L. Rudnick and the anonymous referee for their comments on
the manuscript. The National Radio Astronomy Observatory is a facility
of the National Science Foundation operated under cooperative
agreement by Associated Universities, Inc.

{}

\end{document}